\documentclass[conference]{IEEEtran}
\IEEEoverridecommandlockouts
\usepackage[T1]{fontenc}
\usepackage[utf8]{inputenc}
\usepackage{cite}
\usepackage{amsmath,amssymb,amsfonts}
\usepackage{graphicx}
\usepackage{textcomp}
\usepackage{xcolor}
\usepackage[hyphens]{url}
\usepackage{microtype}
\usepackage[hidelinks]{hyperref}
\begin{document}

\title{Multimodal Takeover Requests for Drivers with Hearing Loss: Implications for AI-Enabled Communication in Automated Vehicles%
\thanks{This work was supported by the National Science Foundation (grant number 2153504).}}

\author{%
\IEEEauthorblockN{Aries Chu, Wei-Hsiang Lo, and Gaojian Huang}
\IEEEauthorblockA{Department of Industrial and Systems Engineering, San Jose State University, San Jose, CA 95192, USA\\}
}

\maketitle

\begin{abstract}
More than 430 million people worldwide live with disabling hearing loss. Although people with hearing loss are legally permitted to drive and may benefit from conditionally automated vehicles, SAE Level 3 systems still require drivers to respond to takeover requests when automation reaches its limits. Existing takeover requests often rely on auditory information, yet little evidence addresses visual and tactile designs for drivers who cannot rely on sound. This driving-simulator study with 40 participants examined the effects of information type (instructional, informative, and baseline), signal type (visual, tactile, and visual-tactile), and hearing condition (normal hearing and simulated hearing impairment) on takeover performance. Information type significantly affected reaction time, with baseline displays producing the shortest times. Signal type significantly affected reaction and takeover time, with visual-tactile displays producing the shortest times. The interaction between signal type and information type was significant for all three measures. Visual-tactile displays produced the shortest reaction times within every information type. With visual-tactile signaling, simple baseline alerts prompted the fastest reactions and the most abrupt maneuvers, whereas informative content produced the lowest mean maximum resulting acceleration. Hearing condition showed no significant main effect on any measure. These findings suggest that AI-enabled vehicles can support urgent takeover communication through visual-tactile displays and can adapt message content to the time available and the maneuver quality required, with implications for drivers across hearing abilities.
\end{abstract}


\vspace{0.8em}
\noindent\fbox{%
\begin{minipage}{0.96\linewidth}
\small
\textbf{Preprint Note.} This manuscript is a preprint and has been submitted for peer review. The content may be revised before formal publication.
\end{minipage}}
\vspace{0.8em}

\section{Introduction}

Hearing loss refers to reduced auditory function and ranges from mild
difficulty to profound loss, including deafness \cite{FloresRamones2023}. More than 430 million people worldwide experience
disabling hearing loss. The World Health Organization~\cite{WHO2024} projects
that this number will exceed 700 million by 2050, meaning that
approximately one in every 10 people will have disabling hearing loss.
Hearing loss is also common among drivers. In a cross-sectional study of
65,533 heavy-vehicle drivers conducted between 2006 and 2016, 26.8\% had
hearing loss, with hearing loss being more prevalent and more severe in
the left ear than in the right ear \cite{Pourabdian2019}. These
prevalence patterns highlight the need to consider hearing access in
vehicle safety design.

People who are deaf or hard of hearing are licensed to drive throughout
the United States, yet reduced access to auditory warnings may affect
awareness of events outside the visual field \cite{NeesWalker2011}.
Edwards et al.~\cite{Edwards2017} followed 500 older adults over three years and
found that moderate or greater hearing loss was associated with poorer
Useful Field of View performance, an indicator of increased risk for
adverse driving events. Hearing level was not significantly associated
with changes in driving mobility or driving cessation. In an on-road
study of 107 community-living older adults, participants with
moderate-to-severe hearing loss performed more poorly in the presence of
distractors than participants with normal or mild hearing loss \cite{Hickson2010}. These findings support vehicle technologies that make
safety-critical information available without relying on sound.

Automated vehicles offer one path toward that support because automation
can reduce the human errors behind many crashes. At SAE Level 3
conditional automation, the automated driving system performs the
dynamic driving task within a defined operational design domain, while
the driver remains receptive to a takeover request \cite{SAE2024}. Automated driving systems can still reach operational limits or
experience failures, requiring the driver to resume manual control
\cite{Danner2023,Dixit2016}. In these situations, drivers
need to rebuild adequate situation awareness within a limited time frame
to prevent a crash \cite{Lu2017}. The takeover process consists of
two phases: the signal response phase and the post-takeover phase \cite{HuangPitts2022}. During the signal response phase, the driver perceives
the takeover request, shifts attention away from any non-driving
activity and back to the road, and prepares to act. During the
post-takeover phase, the driver resumes manual control and performs the
maneuvers the situation requires, such as braking or steering around a
hazard, until the automated system can handle normal driving again.
Previous research suggests a critical window of five to eight seconds
for drivers to regain control, with seven seconds often providing an
appropriate takeover lead time \cite{ErikssonStanton2017,Mok2017}. Within this short window, an effective takeover request informs
the driver when a takeover becomes needed \cite{Zeeb2015} and shapes
driver behavior throughout the takeover process \cite{HuangPitts2022,LoHuang2024}. The safety of drivers with hearing loss in automated
vehicles depends in part on takeover communication that clearly signals
the need for driver intervention \cite{Shull2022}.

In practice, the human-machine interface (HMI) delivers takeover
requests through displays that reach the driver through one or more
sensory channels. These cues should be easy to notice so that the driver
can grasp an urgent traffic situation without extra mental effort, and
designing cues that meet this standard remains difficult \cite{Pececnik2023}. Prior studies have reported benefits from multimodal
warnings that combine visual (V), auditory (A), and/or tactile (T) cues,
and these warnings have been associated with shorter braking times and
higher perceived urgency \cite{Biondi2019,Geitner2019,Liu2023}. Visual displays can offer clear instructions and mark
potential road hazards. For example, a red triangle lighting up in the
side mirror can warn the driver about a vehicle in the blind spot during
a lane change, and many visual takeover cues use red, attention-grabbing
icons \cite{SalubreNathan2021}. Auditory alerts, such as alarms
or voice prompts, can signal the need to take over even when the driver
is looking away from the driving environment. However, the auditory
channel has practical limits. Drivers may disable auditory alarms that
they find annoying \cite{MohdZaki2021}, and drivers with hearing
loss may not perceive these alerts reliably. Tactile feedback can convey
directional information through vibrotactile patterns, such as a
vibrating driver's seat that signals the driver to take over \cite{HuangPitts2022,Petermeijer2017}. As a result, drivers with
hearing loss may need to depend on well-designed visual and tactile
channels to receive takeover requests.

Extensive research has compared multimodal displays with unimodal
displays for vehicle takeover performance. With unimodal displays,
individual preferences strongly influenced how drivers reacted to
takeover requests \cite{LeeKang2023}. Multimodal displays combine two
or three sensory channels, including visual-auditory (VA),
visual-tactile (VT), auditory-tactile (AT), and visual-auditory-tactile
(VAT) formats, and these combinations have been associated with shorter
takeover times than unimodal displays \cite{Bazilinskyy2017,Petermeijer2017}. Multiple stimuli reaching the driver through
different senses heighten the sense of urgency, which supports quicker
and more accurate responses \cite{Petermeijer2017,vanErpEtAl2015}. In an experiment comparing all seven combinations of V, A, and T
cues, the trimodal VAT combination produced the shortest response times
to takeover alerts \cite{HuangEtAl2019}. Similarly, Yun and Yang~\cite{YunYang2020}
reported faster reactions and shorter lane-change times for VAT warnings
than for VA warnings. The best-performing combinations in this
literature include the auditory channel, so the evidence offers little
guidance for drivers who cannot reliably access it. Whether a VT display
still outperforms the V or T display alone when the auditory channel is
unavailable has received far less attention, and this question motivated
the comparison of V, T, and VT displays in the present study.

Beyond the choice of sensory channels, in-vehicle HMIs can be enhanced
by artificial intelligence (AI). AI-enabled automated vehicles can
interpret the traffic environment and the driver's state and deliver
real-time, context-aware assistance, allowing the HMI to convey more
meaningful content than an abstract alert \cite{Huang2025}. Huang et al.~\cite{Huang2025} described three roles in human-AI collaboration. Advisor AI
supports human decision making by providing information without directly
controlling the vehicle. Co-Pilot AI shares control with the human and
provides dynamic action support that adjusts to human actions. Guardian
AI overrides control when the system determines that the human cannot
handle the situation. The timing of the takeover process makes Advisor
AI the most relevant role when a takeover request is issued. At that
moment, the driver has not yet resumed manual control, leaving no human
action for a Co-Pilot AI to adjust to and no dangerous input for a
Guardian AI to override. The Co-Pilot and Guardian roles may assist
after the driver resumes manual control, whereas Advisor AI communicates
through the takeover request itself. What Advisor AI should communicate
at that moment remains unclear. In particular, whether the request
should inform the driver about the danger, such as its location, or
instruct the driver how to act may shape takeover performance.

The distinction between top-down and bottom-up processing offers one way
to explain how drivers interpret takeover information \cite{Neisser1967}.
In top-down processing, a person starts from a broad perspective, such
as context or goals, and then fills in gaps or infers meaning,
consistent with schema theory \cite{Bartlett1932}. In bottom-up processing,
a person starts from sensory details and combines the raw input into a
higher-level understanding \cite{TreismanGelade1980}. Takeover requests
have been designed along these same two routes \cite{HuangPitts2023}.
Informative displays may engage more top-down interpretation because
they describe the driving environment, such as the location and status
of nearby vehicles or pedestrians, and require drivers to choose an
action based on prior experience with these situations (e.g., \cite{HuangPitts2023,MartinezHuang2024}). This
information type has been associated with better situation awareness
because it offers an overall view of the driving environment
\cite{CohenLazry2018,Endsley2020}, and it promotes smoother
transitions when drivers anticipate potential events and prepare for a
takeover \cite{Gold2013}. Instructional displays reduce this
interpretive step by mapping the cue directly to a maneuver. They
present maneuver suggestions, such as whether to turn left or right
along with cues for acceleration or deceleration, and drivers can act on
the suggestion before forming the full picture of the situation (e.g., \cite{LoHuang2024,Meng2015,Petermeijer2017}). The
clear and unambiguous content of instructional displays has been
associated with faster reaction times \cite{VanErpVanVeen2004}, and
it can guide driver attention in complex situations.

Empirical comparisons of the two information types have produced mixed
results. Huang and Pitts~\cite{HuangPitts2023} compared instructional and informative
takeover signals with 24 participants in an SAE Level 3 driving
simulator and found that instructional signals produced marginally
shorter response times than informative signals. In contrast, Martinez and Huang~\cite{MartinezHuang2024} tested display type, information type, and age group
with 21 younger and older adults in a medium-fidelity driving simulator
and found a marginally significant difference in the opposite direction,
with informative displays associated with faster processing than
instructional displays. The choice between the two information types
matters because the takeover window leaves only seconds for the driver
to decode the message and prepare the maneuver \cite{ErikssonStanton2017}. Content that speeds the initial reaction may leave the driver
less prepared for the coming maneuver, whereas content that supports
preparation may cost time up front \cite{Gold2013}. The stakes rise
for drivers with hearing loss, whose visual and tactile channels have to
convey the alert and its content at the same time. The mixed findings
may also reflect the modality used to deliver the content. The same
message may be easy to interpret from a familiar icon but difficult to
decode from a novel vibration pattern. Few studies have tested whether
the effect of information type depends on the signal type that delivers
it \cite{Merat2014}. A clearer understanding of these effects would
allow designers to build takeover requests around the time and
preparation each situation allows.

Despite these findings, evidence remains limited on visual and tactile
takeover cues when auditory access is reduced, especially in SAE Level 3
simulated-driving scenarios. Prior comparisons of takeover displays and
information types have drawn almost entirely on drivers with normal
hearing and full access to sound. Closing this gap matters for two
reasons. Automated vehicles could expand safe and independent mobility
for people with hearing loss, and evidence on their perception of and
responses to visual and tactile cues would help designers create warning
systems that work when sound cannot deliver the message. The present
study takes a controlled first step by comparing visual and tactile
takeover requests under normal-hearing and simulated hearing-impairment
conditions.

The goal of this study was to examine the effects of signal type (V, T,
and VT), information type (instructional, informative, and baseline),
and hearing condition (normal hearing and simulated hearing impairment)
on takeover performance. Because the displays used only visual and
tactile channels, competing predictions were plausible for hearing
condition \cite{Wickens2008}. If reduced auditory input raises overall
attentional load or degrades monitoring of the driving scene,
participants in the simulated hearing-impairment condition would respond
more slowly during takeovers. If takeover performance depends only on
the channels the displays occupy, reducing auditory access should leave
performance unchanged. In both hearing conditions, we expected the VT
display to outperform either unimodal display, given the redundancy
benefits reviewed above \cite{Petermeijer2017}. Given the mixed
findings in prior work, we further examined whether instructional or
informative content would better support the takeover process and
whether the effect of information content depended on the signal
modality. The findings offer implications for how AI-enabled vehicles,
and the Advisor AI role in particular, should communicate urgent
takeover information to drivers with different hearing abilities.

\section{Methods}

\subsection{Participants}

A total of 42 participants (26 females, 16 males) were recruited from
San Jose State University's research pool (SONA) system. Ages ranged from 18 to 44
years, and participants drove an average of 11 hours per week. All
participants received two hours of class credit as compensation.
Eligibility criteria included holding a valid driver's license, having
normal or corrected-to-normal vision, and reporting no cognitive or
neurological conditions that could affect tactile perception. Two
participants did not complete the study, and data from the remaining 40
participants (20 per hearing condition) were included in the analyses.
Approval was obtained from the San Jose State University Institutional Review Board before data collection (IRB protocol 24-235).

\subsection{Apparatus and Stimuli}

\begin{figure*}[!t]
\centering
\includegraphics[width=0.88\textwidth]{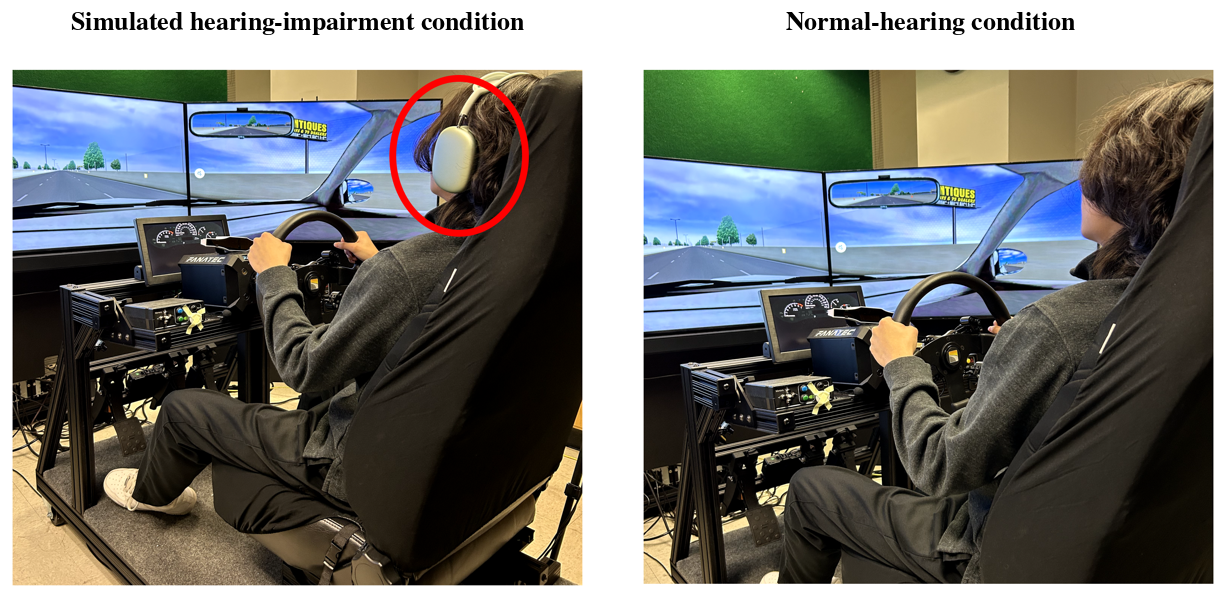}
\caption{The miniSim driving simulator, with a participant completing drives in the simulated hearing-impairment condition (left) and the normal-hearing condition (right).}
\label{fig:minisim}
\end{figure*}

The experiment used miniSim, a medium-fidelity driving simulator
developed by the University of Iowa Driving Safety Research Institute
(Fig.~\ref{fig:minisim}). Three 48-inch LED screens in front of the driver presented
the driving environment, and a 24-inch LCD screen behind the steering
wheel displayed the vehicle odometer. The simulator included a steering
wheel, brake and accelerator pedals, and a driver's seat. A red Start
button turned on the vehicle, and a D indicator showed that the vehicle
was in drive mode. After starting the vehicle, drivers could press a
dedicated button to engage the automated driving system. For tactile
stimulus presentation, nine tactors were installed across the seat back
and seat pan. These tactors delivered vibration stimuli throughout the
experimental scenarios.

\begin{figure*}[!t]
\centering
\includegraphics[width=0.93\textwidth]{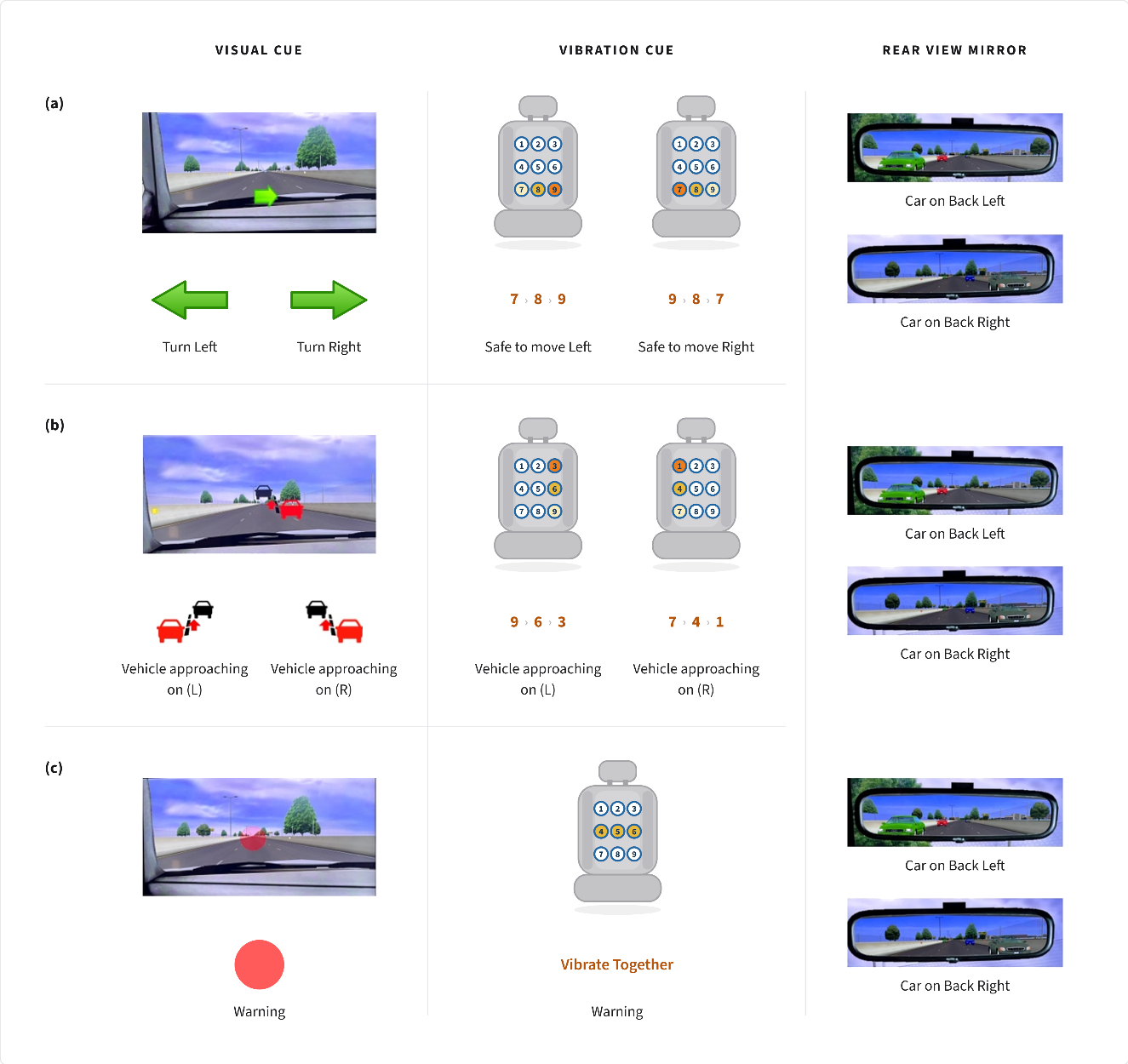}
\caption{Visual and tactile signals in the (a) instructional, (b) informative, and (c) baseline scenarios. Tactile signals were delivered through nine tactors (numbered 1--9) arranged in a $3 \times 3$ array in the driver's seat and activated in the sequences shown.}
\label{fig:signals}
\end{figure*}

Each participant experienced three scenario types: baseline,
instructional, and informative, each characterized by different visual
and tactile patterns (Fig.~\ref{fig:signals}). Visual signals were displayed at 200 $\times$
200 pixels and consisted of a red circle (baseline), a green arrow
(instructional), and a depiction of a red car approaching a black car
(informative). Tactile signals were presented through C-2 tactors (1 $\times$
0.5 $\times$ 0.5 inch; Engineering Acoustics, Inc.) attached to the seat back.
Each tactor vibrated at 250 Hz. In the instructional scenario, visual
navigation signals, such as a left or right turn arrow, guided drivers
toward the safer lane (Fig.~\ref{fig:signals}(a)). For example, a green arrow pointing
left instructed drivers to move to the left lane when another vehicle
occupied the right lane, and the corresponding tactile pattern indicated
the intended maneuver direction. In the informative scenario, the visual
signal depicted a red car approaching from behind a black car on the
left or right side, matching the location of the approaching vehicle
(Fig.~\ref{fig:signals}(b)). The corresponding tactile signal moved vertically along the
right or left side of the seat to convey the side of the approaching
vehicle. In the baseline scenario, signals contained no directional or
environmental content (Fig.~\ref{fig:signals}(c)). The visual signal was a red circle,
and the tactors in the middle row vibrated together, prompting drivers
to check the side mirrors and rearview mirror and select the safest
lane.

The study simulated hearing impairment with participants without
reported hearing loss instead of recruiting drivers with hearing loss.
Simulation held the degree and onset of auditory attenuation constant
across participants and allowed random assignment to hearing conditions,
providing a controlled first step before studies with the target
population. Participants assigned to the simulated hearing-impairment
condition wore earplugs beneath noise-canceling headphones that played
white noise, reducing access to ambient auditory input during the drives
(Fig.~\ref{fig:minisim}). Participants in the normal-hearing condition completed the
same drives without this equipment, allowing the study to examine
whether reduced auditory access affected responses to visual and tactile
cues.

\subsection{Experimental Design}

The study employed a 3 $\times$ 3 $\times$ 2 mixed factorial design. Information type
(baseline, instructional, and informative) and signal type (visual
{[}V{]}, tactile {[}T{]}, and visual-tactile {[}VT{]}) were
within-subject factors, and hearing condition (normal hearing and
simulated hearing impairment) was a between-subject factor. Each
participant completed three drives, one per information type, and each
drive contained six takeover events in which the signal types appeared
in randomized order.

\subsection{Procedure}

Before the experiment, participants completed a questionnaire about
their background and driving experience and signed a consent form. They
were then seated in the driving simulator, which supported both manual
and automated (SAE Level 3) control. Participants completed a 10-minute
training session to become familiar with the simulator and the visual
and tactile cues used in the scenarios. Participants also practiced a
non-driving-related task in which they identified the odd one out among
four options displayed on the left or right side of the screen and spoke
their choices aloud for the researchers to record. The task followed the
procedure used by Huang and Pitts~\cite{HuangPitts2022}.

The training also included a shortened drive that introduced the
scenario elements, including a takeover event and the expected actions.
In this segment, the participant's vehicle began in the middle lane with
automated driving engaged at 60 mph, as in Huang and Pitts~\cite{HuangPitts2022}. A
lead vehicle traveled at a 4-second time headway in the middle lane, and
four other vehicles trailed the lead vehicle at 60 mph, two on each side
of the participant's vehicle (similar to \cite{He2021}). A
construction zone then appeared ahead and blocked the middle lane. The
lead vehicle came to a complete stop at each construction zone with an
approximately 4- to 7-second lead time, and takeover signals (visual,
tactile, or both) prompted the driver to assume control. Participants
were instructed to respond promptly by first braking to deactivate the
automation and then steering into the right or left lane to bypass the
construction zone. They were asked to maintain a speed of 60 mph and to
return to the middle lane once the obstruction was clear. The choice of
lane depended on the distance of the vehicles in the adjacent lanes, and
participants had to avoid the obstacle as well as a possible rear-end
collision with the trailing vehicles. After the simulated drive, the
screen faded to black. Participants then proceeded to the experimental
drives. The researcher briefed them on the drive specifics before each
drive began, and participants assigned to the simulated
hearing-impairment condition wore the earplugs and noise-canceling
headphones with white noise described above. During each 15-minute
drive, a lead vehicle maintained a 7-second time headway in front of the
participant's vehicle. Each drive included six takeover events presented
in randomized order, with signal types also randomized. After each of
the three drives, participants had a five-minute break. After all three
drives were completed, participants were debriefed and compensated. The
full session took about 90 minutes.

\subsection{Dependent Measures}

Dependent measures included reaction time, takeover time, and maximum
resulting acceleration, which served as an indicator of takeover
quality. Reaction time is the interval between the onset of a takeover
signal and the driver's first response to the situation \cite{McDonald2019}. In this experiment, reaction time was measured from signal
onset to the driver's first input, defined as either a steering-wheel
movement or a brake-pedal press. Takeover time was defined as the
interval from signal onset to a 2$^{\circ}$ change in steering-wheel angle,
indicating initiation of the evasive steering maneuver \cite{Louw2017,WanWu2018}. Maximum resulting acceleration was calculated as
the largest vector magnitude of longitudinal and lateral acceleration,
in meters per second squared (m/s$^{2}$), during the post-takeover phase
\cite{Wandtner2018}. This measure reflects takeover quality. Higher
values indicate a more abrupt transition from automated to manual
control, whereas lower values indicate a smoother takeover.

\subsection{Data Analysis}

Separate 3 $\times$ 3 $\times$ 2 mixed analyses of variance (ANOVAs) examined the
effects of information type, signal type, and hearing condition on each
dependent measure. Information type and signal type were within-subject
factors, and hearing condition was a between-subject factor. Degrees of
freedom were corrected using the Greenhouse-Geisser adjustment when the
sphericity assumption was violated, and partial eta squared
($\eta_p^2$) values were reported as effect sizes. The
significance level was .05 for all tests. Significant effects were
followed by Bonferroni-corrected pairwise comparisons, and significant
interactions were examined with simple-effects analyses using
Bonferroni-corrected comparisons. Analyses were conducted in IBM SPSS
Statistics.

\section{Results}

\subsection{Reaction Time}

\begin{figure}[!t]
\centering
\includegraphics[width=\columnwidth]{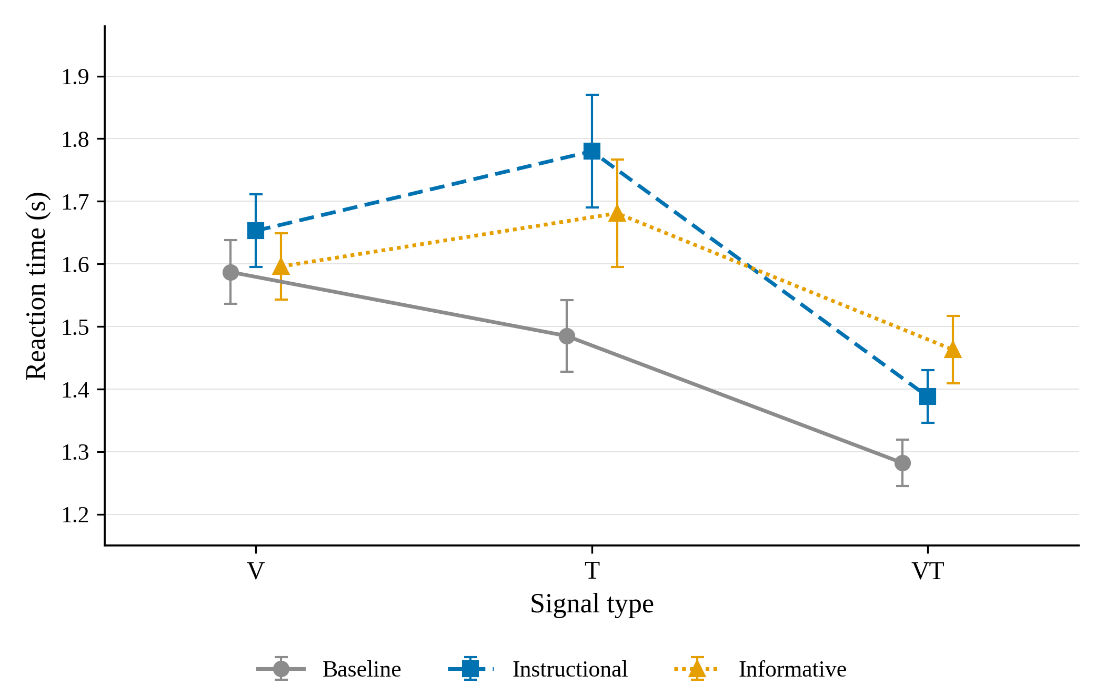}
\caption{Mean reaction times by signal type and information type. Error bars represent standard errors of the mean.}
\label{fig:reaction}
\end{figure}

There were significant main effects of signal type, \emph{F}(1.579,
59.996) = 24.697, \emph{p} $<$ .001, $\eta_p^2$ = .394,
and information type, \emph{F}(1.978, 75.163) = 7.768, \emph{p}
$<$ .001, $\eta_p^2$ = .170, on reaction time (Fig.~\ref{fig:reaction}). For signal type, Bonferroni-corrected pairwise comparisons indicated
that reaction times were shorter for the VT display (\emph{M} = 1.378 s,
\emph{SEM} = 0.037) than for the V (\emph{M} = 1.612 s, SEM = 0.045) and
T (\emph{M} = 1.648 s, SEM = 0.063) displays (\emph{ps} $<$
.001), with no significant difference between the V and T displays
(\emph{p} $>$ .999). For information type, pairwise
comparisons indicated that reaction times were shorter for the baseline
display (\emph{M} = 1.451 s, \emph{SEM} = 0.040) than for the
instructional (\emph{M} = 1.607 s, \emph{SEM} = 0.049, \emph{p} = .001)
and informative (\emph{M} = 1.580 s, \emph{SEM} = 0.058, \emph{p} =
.014) displays, with no significant difference between the instructional
and informative displays (\emph{p} $>$ .999). The main effect
of hearing condition on reaction time was not significant, \emph{F}(1,
38) = 2.433, \emph{p} = .127, $\eta_p^2$ = .060. The
interaction between signal type and information type was significant,
\emph{F}(2.644, 100.481) = 3.728, \emph{p} = .017, $\eta_p^2$ =
.089. Simple-effects analyses showed that reaction times for the VT
display were shorter than for the V and T displays within every
information type (all \emph{ps} $\leq$ .004), and the V and T displays did
not differ within any information type (\emph{ps} $\geq$ .196). Information
type had no significant effect within the V display, \emph{F}(2, 37) =
0.983, \emph{p} = .384, $\eta_p^2$ = .050. Within the T
display, \emph{F}(2, 37) = 7.687, \emph{p} = .002, $\eta_p^2$ =
.294, reaction times were shorter for baseline content than for
instructional (\emph{p} = .004) and informative (\emph{p} = .015)
content. Within the VT display, \emph{F}(2, 37) = 8.203, \emph{p} =
.001, $\eta_p^2$ = .307, reaction times were shorter for
baseline content than for informative content (\emph{p} $<$
.001). The interactions between signal type and hearing condition,
\emph{F}(1.579, 59.996) = 1.044, \emph{p} = .344, $\eta_p^2$ =
.027, and between information type and hearing condition,
\emph{F}(1.978, 75.163) = 2.104, \emph{p} = .130, $\eta_p^2$ =
.052, were not significant.

\subsection{Takeover Time}

\begin{figure}[!t]
\centering
\includegraphics[width=\columnwidth]{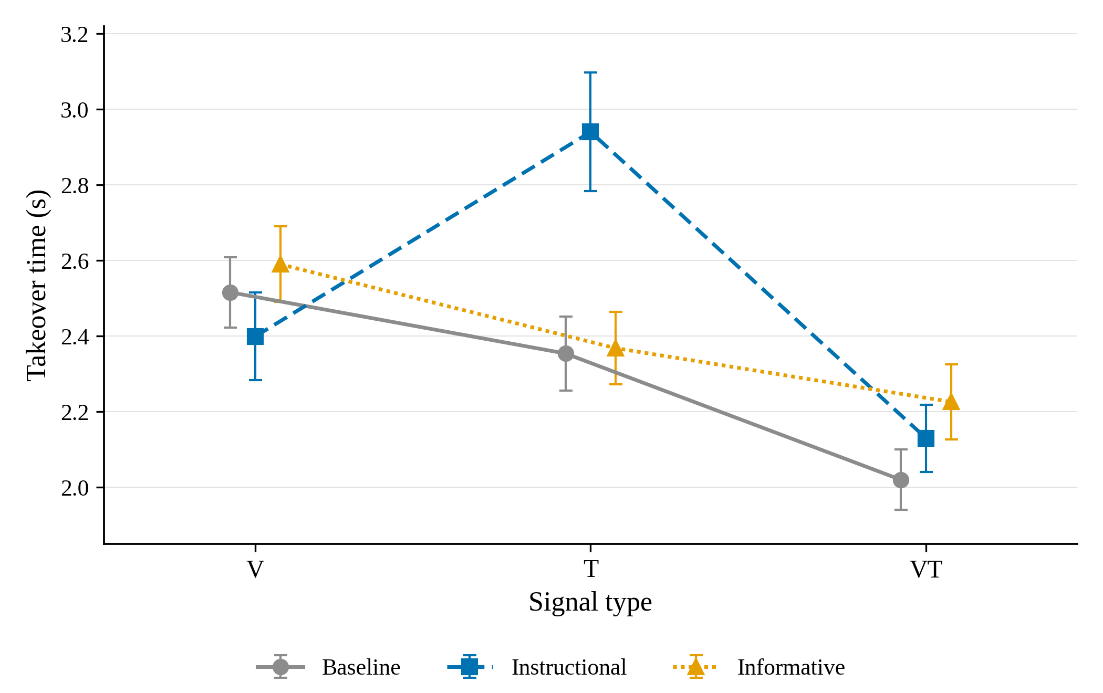}
\caption{Mean takeover times by signal type and information type. Error bars represent standard errors of the mean.}
\label{fig:takeover}
\end{figure}

There was a significant main effect of signal type on takeover time,
\emph{F}(1.738, 66.038) = 18.874, \emph{p} $<$ .001,
$\eta_p^2$ = .332 (Fig.~\ref{fig:takeover}). As in the reaction-time analysis,
pairwise comparisons indicated that takeover times were shorter for the
VT display (\emph{M} = 2.125 s, \emph{SEM} = 0.066) than for the V
(\emph{M} = 2.502 s, SEM = 0.078) and T (\emph{M} = 2.554 s, SEM =
0.087) displays (\emph{ps} $<$ .001), with no significant
difference between the V and T displays (\emph{p} $>$ .999).
The main effects of information type, \emph{F}(1.978, 75.168) = 2.468,
\emph{p} = .092, $\eta_p^2$ = .061, and hearing condition,
\emph{F}(1, 38) = 0.526, \emph{p} = .473, $\eta_p^2$ = .014, on
takeover time were not significant. The interaction between signal type
and information type was significant, \emph{F}(3.253, 123.626) = 8.875,
\emph{p} $<$ .001, $\eta_p^2$ = .189. The takeover-time
cost of the T display was concentrated in the instructional condition.
With instructional content, takeover times for the T display (\emph{M} =
2.940 s, \emph{SEM} = 0.157) were longer than for the V (\emph{p} =
.007) and VT (\emph{p} $<$ .001) displays, and the T display was
slower with instructional content than with baseline (\emph{p} = .001)
or informative (\emph{p} = .005) content. Takeover times for the VT
display were shorter than for the V display within every information
type (\emph{ps} $\leq$ .012) and shorter than for the T display within the
baseline (\emph{p} = .013) and instructional (\emph{p} $<$ .001)
conditions, with no significant difference between the VT and T displays
in the informative condition (\emph{p} = .473). The interactions between
signal type and hearing condition, \emph{F}(1.738, 66.038) = 0.136,
\emph{p} = .845, $\eta_p^2$ = .004, and between information
type and hearing condition, \emph{F}(1.978, 75.168) = 0.740, \emph{p} =
.479, $\eta_p^2$ = .019, were not significant.

\subsection{Maximum Resulting Acceleration}

\begin{figure}[!t]
\centering
\includegraphics[width=\columnwidth]{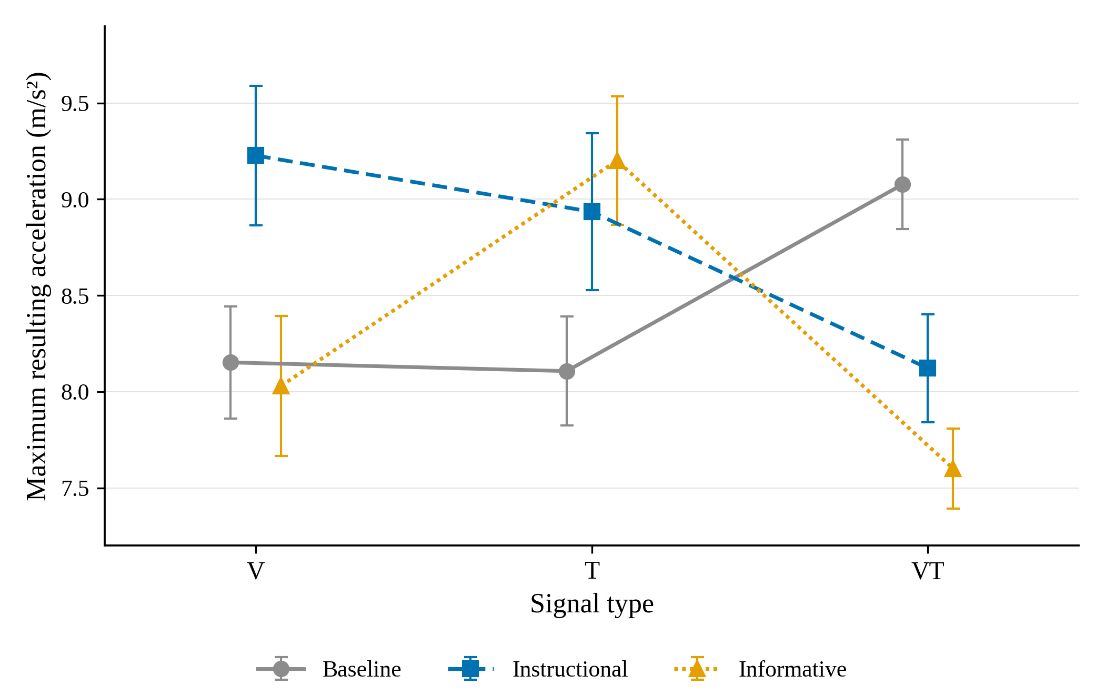}
\caption{Mean maximum resulting acceleration by signal type and information type. Error bars represent standard errors of the mean.}
\label{fig:acceleration}
\end{figure}

No significant main effects of signal type, \emph{F}(1.994, 75.759) =
2.589, \emph{p} = .082, $\eta_p^2$ = .064, information type,
\emph{F}(1.962, 74.551) = 2.603, \emph{p} = .082, $\eta_p^2$ =
.064, or hearing condition, \emph{F}(1, 38) = 1.004, \emph{p} = .323,
$\eta_p^2$ = .026, on maximum resulting acceleration were
found. The interaction between signal type and information type was
significant, \emph{F}(3.714, 141.127) = 10.550, \emph{p} $<$
.001, $\eta_p^2$ = .217 (Fig.~\ref{fig:acceleration}). With the baseline display,
maximum resulting acceleration was higher for the VT signal (\emph{M} =
9.077 m/s$^{2}$, \emph{SEM} = 0.232) than for the V (\emph{M} = 8.152 m/s$^{2}$,
\emph{SEM} = 0.292, \emph{p} = .011) and T (\emph{M} = 8.107 m/s$^{2}$,
\emph{SEM} = 0.283, \emph{p} = .007) signals. With instructional
content, acceleration was higher for the V signal than for the VT signal
(\emph{p} = .006), and with informative content, acceleration was higher
for the T signal than for the V (\emph{p} = .002) and VT (\emph{p}
$<$ .001) signals. Within the VT display, acceleration was
highest with baseline content and lower with instructional (\emph{M} =
8.122 m/s$^{2}$, \emph{SEM} = 0.280, \emph{p} = .010) and informative
(\emph{M} = 7.601 m/s$^{2}$, \emph{SEM} = 0.208, \emph{p} $<$ .001)
content. The interactions between signal type and hearing condition,
\emph{F}(1.994, 75.759) = 1.052, \emph{p} = .354, $\eta_p^2$ =
.027, and between information type and hearing condition,
\emph{F}(1.962, 74.551) = 0.860, \emph{p} = .425, $\eta_p^2$ =
.022, were not significant.

\section{Discussion}

\subsection{Effects of Information Content}

This study examined takeover performance in automated vehicles as a
function of information type, signal type, and hearing condition. For
information type, reaction times were shortest with the baseline
display, and the instructional and informative displays did not differ
reliably from each other. This pattern contrasts with earlier findings
in which meaningful displays supported takeover performance \cite{HuangPitts2023} and in which baseline displays were associated with the
longest processing times \cite{MartinezHuang2024}. One explanation
involves processing depth. Although developed to explain memory, the
levels-of-processing framework suggests that the additional semantic
analysis required by meaningful messages may extend initial processing
time \cite{CraikLockhart1972}. Instructional and informative displays
both contain semantic content that drivers need to decode, either an
action to follow or an environmental state to interpret, whereas the
baseline display works as a simple alarm that requires less semantic
decoding before the driver can act. In a time-critical takeover, this
decoding cost may appear as a longer reaction time even when the added
meaning could support better decisions later in the maneuver. A second
explanation involves familiarity. People respond more readily to stimuli
they have encountered often \cite{MorelandZajonc1982}, and a simple red
warning icon resembles the alerts drivers already know from conventional
vehicles, such as a side mirror that lights up when another car passes.
The baseline display may have benefited from this familiarity, whereas
the instructional and informative patterns were new to participants and
required learning within a short experiment. The reaction-time advantage
of baseline content did not appear as a main effect on takeover time or
acceleration, although the interaction results show that its
consequences depended on signal type.

\subsection{Effects of Visual-Tactile Signaling}

For signal type, the VT display was associated with shorter reaction and
takeover times than either unimodal display, consistent with prior work
showing that multimodal displays support driver alerting \cite{Kaufman2024,LeeEtAl2023}. Multiple resource theory offers a
mechanism for this advantage \cite{Wickens2008}. The theory holds that
human attention draws on separate pools of resources for different
sensory modalities, so information distributed across vision and touch
can be processed in parallel with less interference than information
loaded onto a single channel \cite{Sarter2006,Wickens2008,Yang2022}. A redundant signal in two modalities also raises the chance that
at least one channel captures attention when the driver is looking away
or otherwise occupied. When a takeover request relies on a single visual
or tactile cue, processing is confined to one channel, and the request
competes directly with whatever else occupies that channel. These
findings support multimodal displays as a design guideline for takeover
requests in automated vehicles. Notably, the advantage emerged from
visual and tactile channels alone. Earlier multimodal benefits often
depended on combinations that included sound \cite{HuangEtAl2019,YunYang2020}, and the present results show that a redundancy benefit can
occur without the auditory channel. The simple-effects analyses
strengthen this guideline for the earliest stage of the takeover, since
the reaction-time advantage of the VT display held within every
information type.

\subsection{Signal-Content Trade-Offs Across Takeover Stages}

The interaction between signal type and information type qualified the
main-effect patterns. For reaction time, the advantage of baseline
content appeared with the T and VT displays and was absent with the V
display. Decoding meaning from tactile patterns may have been
particularly demanding because the directional vibration sequences were
new to participants, whereas the visual icons resembled familiar
dashboard symbols. Later in the takeover, the VT and T displays did not
differ significantly under informative content, while the T display
produced markedly longer takeover times with instructional content.
Drivers may have had difficulty translating directional vibration
sequences into action without visual support. The maximum resulting
acceleration findings reveal a related trade-off between response speed
and maneuver smoothness. With the VT display, baseline alerts prompted
the fastest reactions and the most abrupt maneuvers, whereas informative
content slowed the initial reaction and produced the lowest mean maximum
resulting acceleration. This pattern is consistent with a broader
preparation account of takeover performance \cite{Gold2013}. Content
that describes the driving environment may help drivers plan the
maneuver before executing it, so the additional processing time may have
supported smoother control. The interaction patterns indicate that no
single takeover request was optimal across all stages of the transition.
A simple multimodal alert can prioritize detection when time is scarce,
whereas informative content can support smoother control when the
available lead time permits additional processing. This interpretation
extends the two-phase takeover framework \cite{HuangPitts2022} by
showing that signal modality and information content contribute
differently during the signal-response and post-takeover phases.

\subsection{Effects of Hearing Impairment}

Hearing condition produced no significant main effects, a result
consistent with the resource-based prediction outlined in the
Introduction. Auditory information forms a meaningful part of the
driving environment, and background sounds, such as sirens and engine
noise, can inform drivers about events outside their field of view \cite{NeesWalker2011}. Nevertheless, performance in the simulated
hearing-impairment condition did not differ significantly from
performance in the normal-hearing condition on the reported takeover
measures. This finding differs from reports of poorer on-road driving
performance among drivers with hearing loss \cite{Edwards2017,Hickson2010}. Two explanations seem plausible. First, multiple
resource theory implies that reducing auditory input does not directly
reduce the capacity of the visual and tactile channels \cite{Wickens2008}.
Because every takeover request in this study was delivered visually,
tactilely, or both, the auditory channel conveyed no task-relevant
information, and both groups had full access to the channels that
mattered. This account is especially plausible for the present sample,
in which participants did not report hearing loss outside the
experimental manipulation. Second, takeover events demand rapid
responses to sudden and spatially localized hazards. Drivers may rely
mainly on visual and tactile input in these moments, while audition
serves a background monitoring role that can be masked by music, phone
conversations, or road noise \cite{HuangPitts2022,Yang2022}.
Reduced auditory access may consequently have mattered little for the
specific task studied here, although it may matter more in environments
rich in auditory cues, such as urban streets with sirens and horns.
Because the analyses tested for differences instead of statistical
equivalence, the nonsignificant effects should not be interpreted as
evidence that the two hearing conditions performed equivalently. With 20
participants per condition and a temporary simulation of hearing
impairment, the study may not have detected smaller between-condition
differences.

\subsection{Implications for AI-Enabled Takeover Communication}

The findings have implications for AI-enabled vehicles in which an
Advisor AI communicates urgent information to the driver \cite{Huang2025}. The interaction results indicate that an Advisor AI should treat
signal modality and message content as a joint design decision. The
reaction-time advantage of the baseline display suggests that a simple
and familiar alert can prompt a faster initial response than
semantically rich content when the time budget is short. An Advisor AI
could use a simple alert when seconds matter most and informative
content when a longer lead time allows the driver to prepare the
maneuver. The acceleration results refine this guidance. A simple alarm
bought speed at the cost of abrupt control when paired with multimodal
signaling, whereas informative content produced lower mean acceleration.
A short time window could trigger a simple VT alert, whereas a longer
lead time could support an informative VT message that allows the driver
to prepare the maneuver. Instructional tactile messages warrant
particular caution because the tactile-only instructional condition
produced the longest takeover times. The consistent advantage of the VT
display indicates that an Advisor AI could distribute urgent messages
across visual and tactile channels, a strategy with the potential to
serve drivers with hearing loss. User preferences for AI roles and
display modalities vary across individuals and platforms \cite{LoEtAl2025}, so an Advisor AI could further personalize the channel and
content of takeover requests to each driver's sensory profile.

\subsection{Limitations and Future Work}

Several limitations should be noted. First, some participants reported
difficulty perceiving vibrations delivered through the seat, and the
vibration intensity felt strong to some participants and weak to others.
Future designs should calibrate vibration intensity to individual
sensitivity and explore alternative body locations for tactile cues,
such as the wrist or feet, to improve tactile perception for drivers
with hearing loss. Second, hearing impairment was simulated with
earplugs and noise-canceling headphones worn by participants without
reported hearing loss. Simulated impairment reduces auditory input but
cannot reproduce the long-term perceptual adaptations of people who are
deaf or hard of hearing, such as greater reliance on visual scanning.
The degree of auditory attenuation was not reported, so the simulated
condition cannot be mapped to a clinical severity of hearing loss.
Future studies should recruit drivers with actual hearing loss to
capture their needs and preferences directly. Third, the sample consisted of adults affiliated with San Jose State University, aged 18 to 44 years, and included 20 participants per condition. The findings may not generalize
to older drivers or detect small hearing-condition effects. Fourth,
participants experienced the displays for a short period, and every
takeover request in the study was valid. Real systems produce false and
missed alarms, and signal reliability shapes trust in automation and
long-term reliance \cite{HoffBashir2015,LeeSee2004}, so longer
exposure with imperfect reliability deserves study. Finally, the driving
scenario took place on a three-lane highway. Future research should
examine urban environments, where auditory cues such as sirens and horns
are more common and takeover scenarios are more varied, to test whether
hearing condition matters more in those settings.

\section{Conclusion}

Conditionally automated vehicles are increasingly available, yet drivers
still need to resume control when automation reaches its limits. This
transition remains a human factors challenge for people with hearing
loss, who can drive legally but cannot always rely on the auditory
alerts assumed by many vehicle warning systems. Multimodal displays
offer a promising direction because they can alert drivers through
visual and tactile channels.

In this driving-simulator experiment, we investigated the effects of
information type (instructional, informative, and baseline), signal type
(visual {[}V{]}, tactile {[}T{]}, and visual-tactile {[}VT{]}), and
hearing condition (normal hearing and simulated hearing impairment) on
takeover performance. Baseline displays produced the shortest reaction
times, and the VT display produced the shortest reaction and takeover
times. Signal type and information type also interacted, with simple
alerts under multimodal signaling prompting the fastest but most abrupt
responses and informative content producing the lowest mean maximum
resulting acceleration. Hearing condition showed no significant main
effect on any measure. These results indicate that visual-tactile
displays can support faster responses, while the content of the request
shapes both the initial reaction and the quality of the maneuver that
follows. The findings inform AI-enabled communication by showing that an
Advisor AI could combine accessible sensory channels with messages
matched to the time available for takeover and the control quality the
situation requires.

\end{document}